\documentclass[useAMS,usenatbib]{mn2e} 
\usepackage{graphicx}
\usepackage{epsfig}
\usepackage{amsmath}

\newcommand{\hs}{\hspace{1mm}} 
\newcommand{\apj}{ApJ}
\newcommand{\aap}{A\&A} 
\newcommand{\apjl}{ApJL}
\newcommand{\mnras}{MNRAS} 
\newcommand{\aj}{AJ}
\newcommand{\apjs}{ApJS} 
\newcommand{\nat}{{\it Nature}}
 
\newcommand{\physrep}{Phys. Rep.}

\def\lsim{~\rlap{$<$}{\lower 1.0ex\hbox{$\sim$}}}

\def\gsim{~\rlap{$>$}{\lower 1.0ex\hbox{$\sim$}}}

\title[XRB Constraints on the SFR-L$_{\rm X}$ Relation]{Constraints on the Redshift Evolution of the $L_{\rm X}$-SFR Relation from the Cosmic X-Ray Backgrounds}

\author[Dijkstra et al.]{Mark Dijkstra$^{1}$\thanks{E-mail:dijkstra@mpa-garching.mpg.de}, Marat Gilfanov$^{1,2}$, Abraham Loeb$^{3}$, and Rashid Sunyaev$^{1,2}$\\ 
$^1$Max-Planck Institut fuer Astrophysik, Karl-Schwarzschild-Str. 1, 85741 Garching, Germany\\ 
$^{2}$Space Research Institute of Russian Academy of Sciences, Profsoyuznaya 84/32, 117997 Moscow, Russia\\
$^{3}$Astronomy Department, Harvard University, 60 Garden Street, Cambridge, MA 02138, USA}
\def\LaTeX{L\kern-.36em\raise.3ex\hbox{a}\kern-.15em
    T\kern-.1667em\lower.7ex\hbox{E}\kern-.125emX}

\begin{document}

\date{\today} \pagerange{\pageref{firstpage}--\pageref{lastpage}}
\pubyear{2009}

\maketitle

\label{firstpage}

\begin{abstract}
Observations of local star forming galaxies have revealed a
correlation between the rate at which galaxies form stars and their
X-Ray luminosity. We combine this correlation with the most recent
observational constraints on the integrated star formation rate
density, and find that star forming galaxies account for 5-20\% of the
total soft and hard X-ray backgrounds, where the precise number
depends on the energy band and the assumed average X-ray spectral
energy distribution of the galaxies below $\sim 20$ keV. If we combine
the $L_{\rm X}$-SFR relation with recently derived star formation rate
function, then we find that star forming galaxies whose X-ray flux
falls well (more than a factor of $10$) below the detection thresholds
of the Chandra Deep Fields, can fully account for the unresolved soft
X-ray background, which corresponds to $\sim 6\%$ of its
total. Motivated by this result, we put limits on the allowed redshift
evolution of the parameter $c_{\rm X} \equiv L_{\rm X}$/SFR, and/or
its evolution towards lower and higher star formation rates. If we parametrize
the redshift evolution of $c_{\rm X}\propto (1+z)^b$, then we find
that $b \leq 1.3$ (95\% CL).
On the other hand, the observed X-ray luminosity functions (XLFs) of star forming galaxies indicate that $c_{\rm X}$ may be increasing towards higher redshifts and/or higher star formation rates at levels that are consistent with the X-ray background, but possibly at odds with the locally observed $L_{\rm X}$-SFR relation.
\end{abstract}

\begin{keywords}
galaxies: high redshift -- galaxies: stellar content -- X-rays: binaries -- X-rays: galaxies
\end{keywords}
 
\section{Introduction}
\label{sec:intro}

Galaxies contain various sources of X-ray emission which include: ({\it i}) active galactic nuclei (AGN), which are powered by accretion of gas onto a supermassive black hole ({\it ii}) hot ($T\gsim 10^6$ K) interstellar gas, and ({\it iii}) X-ray binaries, which consist of a compact object, either a neutron star or a stellar mass black hole, and a companion star from which the compact object accretes mass. The X-Ray luminosity of star forming galaxies without AGN is dominated by so-called high-mass X-ray binaries (HMXBs,
e.g. Grimm et al. 2003), in which the neutron star or stellar mass black hole is
accreting gas from a companion that is more massive than $\sim
5M_{\odot}$. HMXBs are thus tightly linked to massive stars, and since
massive stars are short lived, the combined X-ray luminosity of HMXBs
is expected to be linked to the rate at which stars form
\citep[e.g.][]{HM01}. The X-Ray luminosity of star forming galaxies is
indeed observed to be correlated with the rate at which they are
forming stars \citep{Grimm03,Ranalli03,Gilfanov04,Persic04,Mineo10,Lehmer10,Mineo11,S11}.

This `$L_{\rm X}$-SFR relation' encodes a wealth of information on
various astrophysical processes. These include the initial mass
function (IMF) of stars, the fraction of massive stars that form
in binaries, the mass ratio distribution of binary stars, the
distribution of their separations, the gas metallicity, and the common
envelope efficiency \citep[e.g.][and references
therein]{B02,Mineo11}. Despite its dependence on many astrophysical
processes, the $L_{\rm X}$-SFR relation is observed to hold over $\sim
4$ orders of magnitude \citep{Mineo11}, with a modest scatter
($\sigma=0.4$ dex).

Existing observations have only been able to probe the $L_{\rm X}$-SFR
relation in nearby galaxies. There are observational hints--as well as theoretical expectations--
that $c_{\rm X}$ is higher in low mass galaxies and/or low metallicity
environments \citep{Dray06,Linden10,K11}, which suggests that $c_{\rm X} \equiv {L_{\rm X}}/{{\rm SFR}}$
could have been substantially higher at higher redshifts
\citep[see][for a summary]{Mirabel11}. However, quantitative
constraints on ratio $c_{\rm X}$ at higher
redshifts are virtually non-existent. The main reason is that the
X-ray flux that reaches Earth from individual star forming galaxies
typically falls well below the detection threshold of existing X-ray
telescopes (see \S~\ref{sec:cxb}).

The relation between SFR and $L_{\rm X}$ also plays an important role in determining the thermal history of intergalactic medium at very high redshifts \citep[e.g.][]{Oh01,Ven01}, and strongly affects the 21-cm signal from atomic hydrogen during the dark ages \citep{Fur06,PF07,PL10,Alvarez10}. Current theoretical models of this reheating process explore values for $c_{\rm X}$ that span $\sim 4$ orders of magnitude.

The goal of this paper is to investigate whether it is possible to put
any constraints on the evolution of $c_{\rm x}$ with either redshift
and/or towards high/low star formation rates than probed by existing
observations of individual galaxies, using the cosmic X-Ray backgrounds (CXBs). 

Our paper is organized as follows. In \S~\ref{sec:cxb}, we summarize
recent observational constraints on the levels of the total X-ray
background (XRB, both soft and hard), as well as the fraction of the
XRB that has been resolved in discrete X-ray sources. In
\S~\ref{sec:totsfr} we compute the total contribution of star forming
galaxies to the soft and hard X-Ray backgrounds, and show that this
contribution can be substantial. In \S~\ref{sec:usfr} we show that
star forming galaxies, too faint to be detected as individual
X-ray sources, can fully account for the unresolved portion of the
XRB. In \S~\ref{sec:lxsfr} we put constraints on possible trends in
the $L_{\rm X}$-SFR relation using the unresolved soft XRB. 
There are several additional candidate sources which contribute to the unresolved CXBs (these include for example low mass X-ray binaries and weak AGN, see \S~\ref{sec:conc}). To constrain the evolution of $c_{\rm X}$ we will allow the entire unresolved SXB to be produced by HMXBs (and the hot ISM) in X-ray faint star-forming galaxies. This results in conservative upper limits on possible evolution in $c_{\rm X}$.
Finally,
we conclude in \S~\ref{sec:conc}. The cosmological parameters used
throughout our discussion are
$(\Omega_m,\Omega_{\Lambda},\Omega_b,h)=(0.27,0.73,0.046,0.70)$
\citep{Komatsu09}.

\section{The Cosmic X-Ray Backgrounds (CXB)}
\label{sec:cxb}

\subsection{The Observed Soft X-ray Background (SXB)}
\label{sec:sxb}

The total soft ($E$=1-2 keV) CXB (SXB) amounts to $S_{1-2}=4.6 \pm 0.3 \times 10^{-12}$ erg s$^{-1}$ cm$^{-2}$ deg$^{-2}$ \citep[e.g.][]{HM06}.  \citet{HM06} find an unresolved SXB intensity of $1.04 \pm 0.14\times 10^{-12}$ erg s$^{-1}$ cm$^{-2}$ deg$^{-2}$, after removing all point and extended sources detected in the {\it Chandra Deep Fields} (CDFs). The detection threshold in the {\it Chandra Deep Field-North} (CDF-North) corresponds to $s_{\rm th}\sim 2.4 \times 10^{-17}$ erg s$^{-1}$ cm$^{-2}$ \citep{A03,HM07}. 

\citet{HM07} showed through a stacking analysis that $\sim 70\%$ of the unresolved component is accounted for by sources that are detected with the {\it Hubble Space Telescope}, but not individually as X-ray sources. Their stacking analysis shows that these X-ray undetected sources have an {\it average} X-ray flux that is $\langle s \rangle \sim 0.15-0.30 s_{\rm th}\sim 3.6-7.2 \times 10^{-18}$ erg s$^{-1}$ cm$^{-2}$. \citet{BG} find that the cumulative number of these X-ray undetected HST sources brighter than some of X-ray flux $s$, is well-described by a power-law $N(>s) \propto s^{-\beta}$, where $\beta=1.1^{+0.5}_{-0.3}$. We can then estimate the minimum X-ray flux, $s_{\rm min}$, that the HST-detected sources probed from $\langle s\rangle=\int_{s_{\rm min}}^{s_{\rm th}}dS \frac{dN}{dS}S/\Big{[}\int_{s_{\rm min}}^{s_{\rm th}}dS \frac{dN}{dS}\Big{]}$, and find that $s_{\rm min}=1.2-3.2 \times 10^{-18}$ erg s$^{-1}$ cm$^{-2}$ (for $\beta=1.1$). Throughout, we use the remaining 30\% of the unresolved SXB as a measure of the `true unresolved SXB', which amounts to $3.4 \pm 1.4 \times 10^{-13}$ erg s$^{-1}$ cm$^{-2}$ deg$^{-2}$ \citep{HM07}, down to a minimum flux $s_{\rm min}$. 

\begin{figure*}
\vbox{\centerline{\epsfig{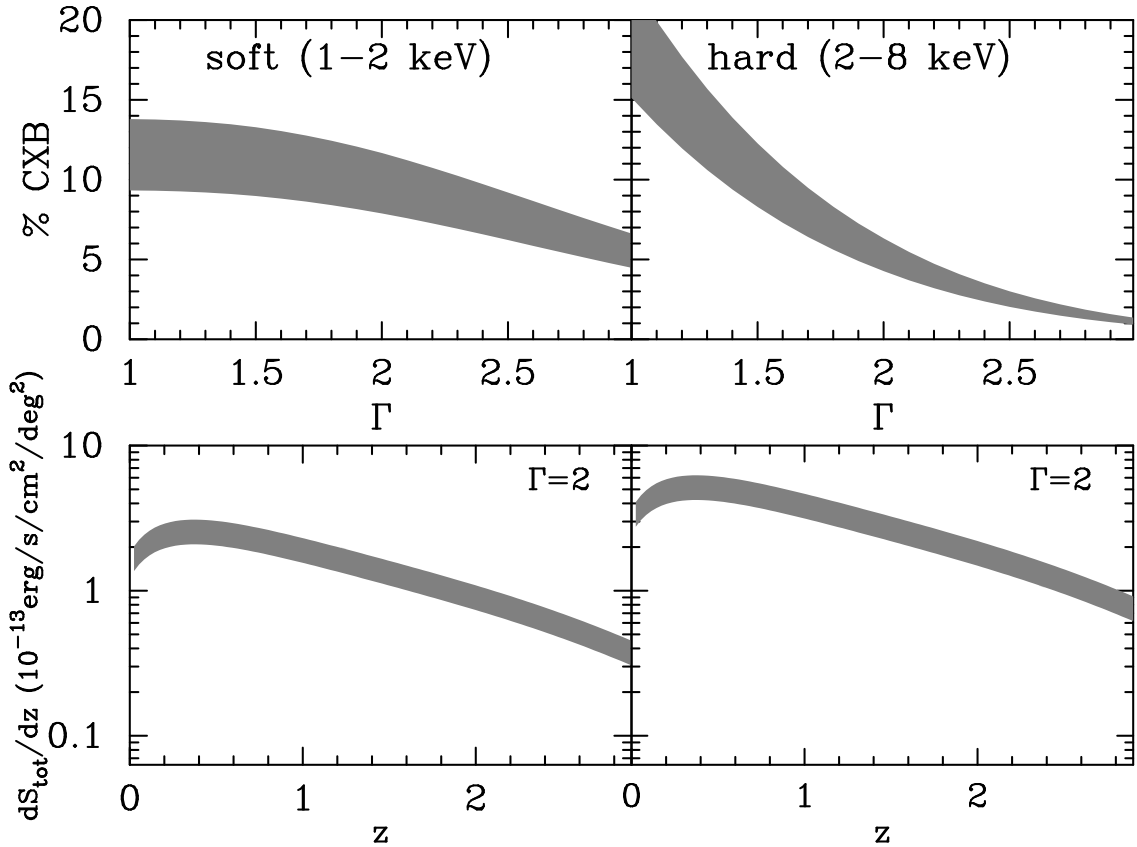}}}
\caption[]{The {\it top panels} show the fraction of the total soft
({\it left}) and hard ({\it right}) X-Ray backgrounds that can be
attributed to star forming galaxies, as a function of the assumed
photon index $\Gamma$, for power-law X-Ray spectral energy
distributions (see text). We have drawn curves for a range of observed
values for $c_{\rm X} \equiv {L_{\rm X}}/{{\rm SFR}}$ (see
text). Depending on the assumed value for $\Gamma$, star forming
galaxies can account for $\sim 5-15\%$ of the total observed soft
X-ray background, and up to $\sim 20\%$ of the hard X-ray
background. For comparison, \citet{Sw04} found that typically ULX
spectra in the Chandra bands could be described by a powerlaw with
$\langle \Gamma \rangle=1.7$, which would place the contribution of
star forming galaxies to the soft X-ray background at $\sim 9-13
\%$. As the spectra of ULXs at $E>10$ keV are poorly known, the contribution of star forming
galaxies to the hard X-ray background is more uncertain. The {\it
lower panels} show that the contribution per unit redshift, $dS_{\rm
tot}/dz$, peaks at low redshift $z\sim 0.4-0.5$. }
\label{fig:fig0}
\end{figure*}
\subsection{The Observed Hard X-ray Background (HXB)}
\label{sec:hxb}
The total hard ($E$=2-8 keV) CXB (HXB) amounts to $S_{2-8}=1.7 \pm 0.2
\times 10^{-11}$ erg s$^{-1}$ cm$^{-2}$ deg$^{-2}$
\citep[e.g.][]{HM06}. \citet{HM06} find an unresolved HXB intensity of
$3.4 \pm 1.7\times 10^{-12}$ erg s$^{-1}$ cm$^{-2}$ deg$^{-2}$. We
will not use the unresolved HXB to put constraints on the redshift
evolution of $c_{\rm X}$ for two reasons: ({\it i}) the uncertainties
on the unresolved HXB are larger than for the SXB, and ({\it ii}) as
we will explain below (\S~\ref{sec:result1}) we are much more
sensitive to the assumed X-Ray spectrum outside the Chandra bands,
when we compare our models to the HXB.

\section{The Contribution of Star Forming Galaxies to the CXB}
\label{sec:totsfr}
\subsection{The Model}
The total contribution $S_{\rm tot}$ (in erg s$^{-1}$ cm$^{-2}$ deg$^{-2}$) of star forming galaxies to the SXB is given by (Appendix~\ref{app:der})

\begin{eqnarray}
\label{eq:stot}
S_{\rm tot}=\frac{\Delta \Omega}{4\pi}\frac{c}{H_0}\int_{0}^{z_{\rm max}}\frac{dz}{(1+z)^2\mathcal{E}(z)} \dot{\rho}_*(z) \mathcal{L}_X(z,\Gamma).
\label{eq:sx}
\end{eqnarray} Here $\Delta \Omega\sim 3.0\times 10^{-4}$ sr deg$^{-2}$, $\mathcal{E}(z)=\sqrt{\Omega_{\rm m}(1+z)^3+\Omega_{\Lambda}}$, and $\dot{\rho}_*(z)$ denotes the comoving star formation rate density at redshift $z$ (in $M_{\odot}$ yr$^{-1}$ cMpc$^{-3}$, where `cMpc' stands for co-moving Mpc). We adopt the star formation history from \citet{Hopkins06}, using the parametric form $\dot{\rho}_*(z)=(a+bz)h/(1+(z/c)^d)$ from \citet{Cole01}, where $a=0.017$, $b=0.13$, $c=3.3$, and $d=5.3$. We note that the amplitude of the function $\dot{\rho}_*(z)$ depends on the assumed IMF \citep[see][for a more detailed discussion]{Hopkins06}. The adopted normalization derives from classical Salpeter IMF \citep{Salpeter55} between 0.1 to 100$M_{\odot}$ \citep{Hopkins04}. The same IMF was assumed in the derivation of the SFR-$L_{\rm X}$ relation \citep[see][]{Mineo11}, and our calculations are therefore self-consistent.

%
The term $\mathcal{L}(z,\Gamma)\equiv c_{\rm X}K(z,\Gamma)$ denotes
the `K-corrected' X-ray luminosity (in erg s$^{-1}$) per unit star
formation rate in the {\it observed} energy range
E$_1$-E$_2$. In this paper, E$_1=1.0$ keV and E$_2=2.0$ keV for the soft band, and E$_1=2.0$ keV and E$_2=8.0$ keV for the hard band.
\citet{Mineo11} recently determined\footnote{The literature contains values for $c_{\rm X}$ that at face value appear both lower \citep[e.g.][]{Persic07} and higher \citep[e.g.][]{Ranalli03} by a factor of a few. However, some studies measured the X-Ray luminosity ($L_{\rm X}$) in the range 2-10 keV \citep[e.g.][]{Gilfanov04,Persic04,Persic07,Lehmer10}, or 0.5-2.0 keV \citep{Ranalli03}. Furthermore, some studies have derived values for $c_{\rm X}$ using a formation rate of stars (SFR) in the mass range $5 \leq M/M_{\odot} \leq100$ \citep[e.g.][]{Persic04}, which results in larger values for $c_{\rm X}$. \citet{Mineo11} discuss that most studies are consistent with their derived value when identical definitions for `SFR' and `$L_{\rm X}$' are used.}
 the value of $c_{\rm X} \equiv L_{\rm X}$/SFR, where SFR denotes the star formation rate in
$M_{\odot}$ yr$^{-1}$, to be $c_{\rm X}=2.6\times 10^{39}$ ${\rm
erg~s^{-1}}/[M_{\odot}~{\rm yr}^{-1}]$ when only compact resolved
X-Ray sources in galaxies are included. \citet{Mineo11} also found that the best fit $c_{\rm X,max}=3.7
\times 10^{39}$ erg s$^{-1}\hs [M_{\odot}\hs{\rm yr}^{-1}]^{-1}$ for
unresolved galaxies in the {\it Chandra Deep Field North} and
ULIRGs. However, a non-negligible fraction of this additional
unresolved flux is in a soft component, and would not contribute to
soft X-Ray background (measured in the 1-2 keV band, see Bogdan \&
Gilfanov 2011). We will not attempt to model in detail the
contribution of unresolved X-Ray emission. Instead,
Figures~\ref{fig:fig0},~\ref{fig:fig1}, and ~\ref{fig:fig2} show our
results for the full range $c_{\rm x}=2.6-3.7 \times 10^{39}$ erg
s$^{-1}\hs [M_{\odot}\hs{\rm yr}^{-1}]^{-1}$. We point out that \citet{Mineo11} measured
the X-ray luminosity over the energy range $0.5-8$
keV.
%

This observed relation between $L_{\rm X}$ and SFR holds over $\sim 4$
orders of magnitude in SFR from SFR$\sim 0.1-1000\hs M_{\odot}$
yr$^{-1}$, with a scatter of $\sigma \sim 0.4$ dex. We include this
scatter in our calculation by convolving equation~(\ref{eq:sx}) with a
lognormal probability distribution function for $c_{\rm X}$ with a
mean of $\langle \log c_{\rm X} \rangle =39.4$ and standard deviation
of $\sigma=0.4$ \citep[see Fig.~9 of][]{Mineo11}. Because the
log-normal distribution is symmetric in $\log c_{\rm X}$, it is skewed
towards larger values of $c_{\rm X}$ in linear coordinates. The
observed scatter in the $L_{\rm X}$-SFR relation therefore enhances
our computed contributions to the X-ray backgrounds. In the absence of
this scatter, $S_{\rm tot}$ reduces by a factor of $\exp\big{(}
{-{1\over 2}\sigma^2 {\rm ln}^2 10}\big{)}\sim 0.65$.

Finally, to compute the contribution of a galaxy at redshift $z$ to the CXB in
the observed energy range E$_1$-E$_2$, we need to compute the galaxy's
luminosity in the range [E$_1$-E$_2$]$\times(1+z)$ keV. In analogy to
the standard `K-correction' \citep[e.g.][]{Hok02}, we multiply $c_{\rm
X}$ by $K_{\rm
X}(z,\Gamma)=\mathcal{I}(E_1(1+z),E_2(1+z))/\mathcal{I}(0.5\hs{\rm
keV},8.0\hs{\rm keV})$. Here, $\mathcal{I}(x,y)=\int_x^y E n(E) dE$,
where $n(E)dE$ denotes the number of emitted of photons in the energy
range $E\pm dE/2$. Note that we assume that the X-Ray SED does not
depend on the star formation rate $\psi$. In this paper, we explore
power law SEDs for the form $n(E) \propto E^{-\Gamma}$, and the
integrals $\mathcal{I}(x,y)$ can be evaluated analytically. Finally,
we take the redshift integral from $z_{\rm min}=0$ to $z_{\rm
max}=10$.  The results are only weakly dependent on the integration
limits we pick, as long as $z_{\rm max} \geq 4.0$ (see the {\it lower
right panel} of Fig.~\ref{fig:fig1}).

\subsection{Results}
\label{sec:result1}

The {\it top panels} of Figure~\ref{fig:fig0} show the fraction of the total soft ({\it left panel}, 1-2 keV) and hard ({\it right panel}, 2-8 keV) X-Ray backgrounds that can be attributed to star forming galaxies, as a function of the assumed photon index $\Gamma$, and for a realistic range of $c_{\rm X}$ (see above). Depending on the assumed value for $\Gamma$, star forming galaxies can account for $\sim 5-15\%$ of the total soft X-ray background, and up to $\sim 20\%$ of the observed hard X-ray background.  For comparison, \citet{Sw04} found that the observed distribution of $\Gamma$ for ultra luminous X-ray sources (ULXs) had a mean $\langle \Gamma \rangle=1.7$, and mode (i.e. most likely value) of $\Gamma_{\rm pk}\sim 2.0$ in the Chandra bands. For the range $\Gamma=1.7-2.0$ the contribution of star forming galaxies to the soft X-ray background is $\sim 8-13 \%$. The contribution to the hard XRB depends more strongly on $\Gamma$ because for this calculation, the K-correction involves a larger extrapolation of the assumed SED, which is poorly known at $E>10$ keV.

The {\it lower panels} show the differential contribution as a function of redshift. This differential contribution is given by
${dS_{\rm tot}}/{dz}$. These plots show that ${dS_{\rm tot}}/{dz}$ peaks at $z\sim 0.38$, and that the dominant contribution to the
total X-Ray background comes from lower redshifts. Indeed, $\sim 50\%$ [$\sim 25\%$] of the total contribution comes from $z \lsim 1.3$ [$z\lsim 0.8$] (see
\S~\ref{sec:resultsusxb}).

\section{Contribution of X-Ray Faint Star Forming Galaxies to the SXB}
\label{sec:usfr}

\subsection{The Model}
\begin{figure*}
\vbox{\centerline{\epsfig{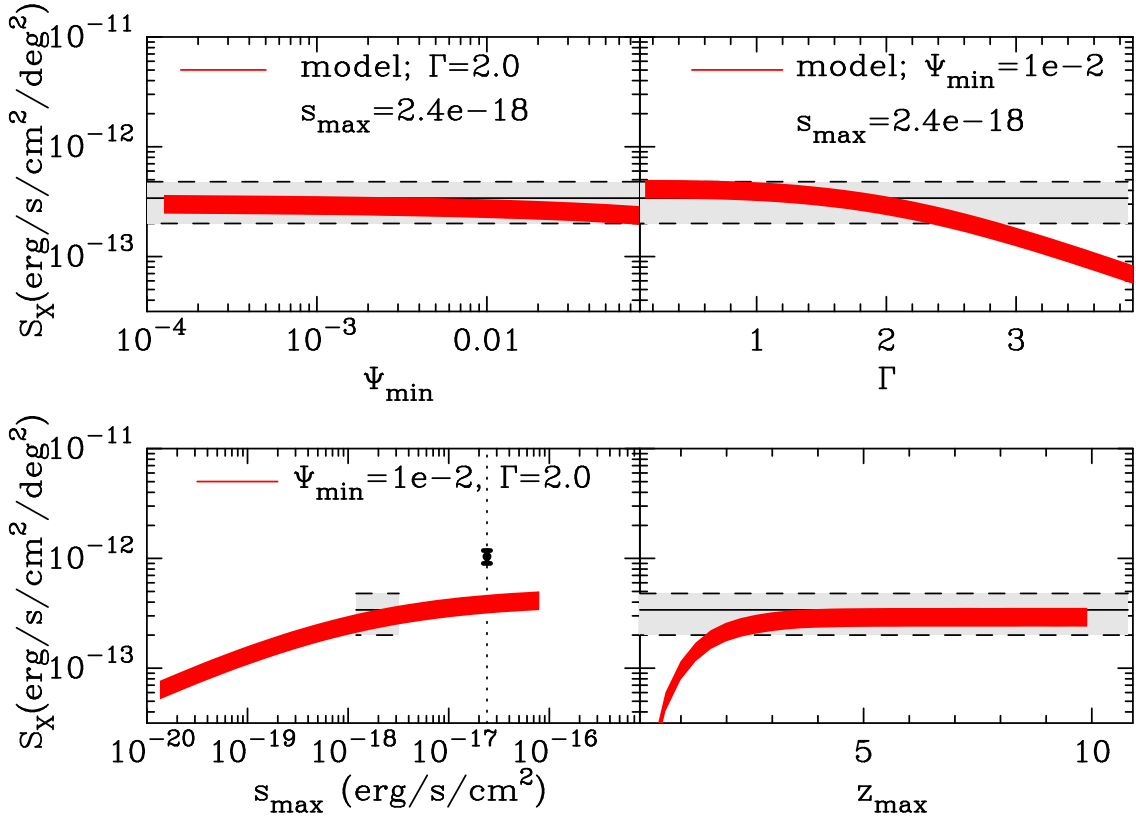}}}
\caption[]{The contribution $S_{\rm X}$ to the soft X-Ray background (SXB, $E$=1-2 keV in the observer's frame) by galaxies whose individual soft X-ray flux is less than $s_{\rm max}$. The median unresolved SXB is represented by the {\it black solid horizontal lines}, and its 68\% confidence levels by the the {\it gray region}, bounded by {\it black dashed lines}  \citep[taken from][]{HM07},. The {\it red solid lines} show $S_{\rm X}$ as a function of various model parameters. Our fiducial model assumes $\Gamma=2.0$, $s_{\rm max}=2.4 \times 10^{-18}$ erg s$^{-1}$ cm$^{-2}$, $\psi_{\rm min}=0.01 M_{\odot}$ yr$^{-1}$, and $z_{\rm max}=10$. The {\it upper left panel} shows that $\psi_{\rm min}$ only weakly affects $S_{\rm X}$, because the faint end of the star formation function (especially at low $z$) is not steep. The {\it upper right panel} shows that $S_{\rm X}$ depends weakly on $\Gamma$, provided that $\Gamma \lsim 2$.  The {\it lower left panel} shows that $S_{\rm x}$ also depends weakly on $s_{\rm max}$, unless $s_{\rm max}\lsim 10^{-18}$ erg s$^{-1}$ cm$^{-2}$. The {\it dotted vertical line} shows the X-ray detection threshold in CDF-N \citep{A03}. The {\it black filled circle} on this {\it line} shows the unresolved SXB derived by Hickox \& Markevitch (2006, i.e. before subtracting the contribution from X-Ray faint HST detected sources).  The {\it gray region} here brackets the effective minimum X-ray flux $s_{\rm min}$ that is probed by stacking X-ray undetected HST sources (see \S~\ref{sec:sxb}). The {\it lower right panel} shows that $S_{\rm X}$ again depends weakly on $z_{\rm max}$, provided that $z_{\rm max}>4$. See the main text for a more detailed interpretation of these plots. These plots show that {\it star forming galaxies that are too faint to be detected as individual X-ray sources, can account for the full unresolved SXB}, and that this statement is insensitive to details in the model when $\Gamma \lsim 2$, which is reasonable given the available observational constraints \citep{Sw04}.}
\label{fig:fig1}
\end{figure*}

We can compute the contribution $S_{\rm X}$  of star forming galaxies, {\it fainter than some observed soft X-ray flux $s_{\rm max}$}, to the SXB as
\begin{eqnarray}
S_X=\frac{\Delta \Omega}{4\pi}\frac{c}{H_0}\int_{0}^{z_{\rm max}}\frac{dz}{(1+z)^2\mathcal{E}(z)} \times \\ \nonumber
\int_{0}^{L_{\rm X, max}} d \log L_{\rm X} \hs n(\log L_{\rm X},z)\hs L_{\rm X}K_{\rm X}(z,\Gamma), 
\label{eq:sxfaint}
\end{eqnarray}  
where $n(\log L_{\rm X},z)d \log L_{\rm X}$ denotes the comoving
number density of star forming galaxies with X-ray luminosities in the
range $\log L_{\rm X} \pm d \log L_{\rm x}/2$ (i.e. the units of
$n(\log L_{\rm X},z)$ are cMpc$^{-3}$ dex$^{-1}$). Here, $L_{\rm X}$
denotes the X-ray luminosity of galaxies in the 0.5-8.0 keV
(restframe). The integral over $L_{\rm X}$ then extends up to $L_{\rm
X,max} \equiv 4 \pi d^2_{\rm L}(z)s_{\rm max}/K_{\rm X}(z,\Gamma)$,
where $d_{\rm L}(z)$ is the luminosity distance to redshift $z$. The
quantity $n(\log L_{\rm X})$, also referred to as the X-ray luminosity
function (XLF) of star forming galaxies, is given by
\begin{eqnarray}
n(\log L_{\rm X},z)=\int_{\psi_{\rm min}}^{\psi_{\rm max}}d\psi\hs n(\psi,z)P(\log L_{\rm X}|\psi),
\label{eq:xlf}
\end{eqnarray}  
where $n(\psi,z)d\psi$ denotes 'star formation rate function', which gives the comoving number density of galaxies that are forming stars at a rate SFR$=\psi \pm d\psi/2$ at redshift $z$. The function $P(\log L_{\rm X}|\psi)d \log L_{\rm x}$ denotes the probability that a galaxy that is forming stars at a rate $\psi$ has an X-ray luminosity in the range $\log L_{\rm X} \pm d \log L_{\rm x}/2$. We describe both functions in more details below. We start the integral over $\psi$ at $\psi_{\rm min}=10^{-3} M_{\odot}$ yr$^{-1}$, which corresponds 
approximately to the SFR that is theoretically expected to occur in dark
matter halos of mass $M_{\rm halo}\sim 10^{8} M_{\odot}$
\citep{Wise,TC07,ZZ1}. Our final results depend only weakly on $\psi_{\rm
min}$ (see the {\it upper left panel} of Fig.~\ref{fig:fig1}). The
$\psi$-integral extends up to $\psi_{\rm max}=10^5 M_{\odot}$
yr$^{-1}$, with our results being insensitive to this choice.

In the local Universe ($z\sim 0$), the function $n(\psi,z)$ [units are cMpc$^{-3}$ $(M_{\odot}\hs{\rm yr}^{-1})^{-1}$] appears\footnote{Previous work showed that this star formation rate function can be described by a log-normal function \citep{Martin05}.  However, this lognormal function does not provide a good fit to the observed star formation rate function particularly for low and high star formation rates (see Fig~4 of Bothwell et al. 2011).} to be described accurately by a Schechter function \citep{Bothwell11}
\begin{equation}
 n(\psi,z)=\frac{\Phi^*}{\psi^*}\Big{(}
 \frac{\psi}{\psi^*}\Big{)}^{\alpha}e^{-\psi/\psi^*},
\end{equation} 
where $\alpha=-1.51\pm0.08$, $\Phi^{*}=(1.6\pm 0.4)\times 10^{-4}$
cMpc$^{-3}$, and $\psi^*=9.2 \pm 0.3 M_{\odot}$ yr$^{-1}$. The
redshift evolution of $n(\psi,z)$ is not well known.  We
assume throughout that $\alpha(z)=-1.51-0.23g(z)$, where $g(x)\equiv
\frac{2}{\pi}{\rm arctan} \hs x$ is a function that obeys $g(0)=0$ and
$\underset{z \rightarrow \infty}{\rm lim}g(z)=1$. This steepening of the low-end of the star formation rate
function at higher redshifts reflects the steepening of UV luminosity
functions towards higher redshifts (e.g. Arnouts et al. 2005, Reddy \&
Steidel 2009, Bouwens et al. 2006, 2007,2008), and the observation
that dust-obscuration is negligible for the UV-faint galaxies
(e.g. Bouwens et al. 2009). The factor `-0.23' causes $\alpha\rightarrow -1.74$ at high redshift, which corresponds to the best-fit slope of the UV-luminosity function of $z=6$ drop-out galaxies (Bouwens et al. 2007).
The redshift evolution of $\Phi^*$ and
$\psi^*$ is more difficult to infer from the redshift evolution of the
UV luminosity functions, because of dust. We have taken two
approaches: we constrain either the redshift evolution of $\Phi^*$ or
$\psi^*$-while keeping the other fixed- to match the inferred redshift
evolution of the star formation rate density (see
Appendix~\ref{app:dndpsi} for more details). In reality, we expect
both parameters to evolve with redshift, and that our two models
bracket the range of plausible more realistic models.

The function $P(\log L_{\rm X}|\psi)$ is given by a lognormal
distribution
\begin{equation}
 P(\log L_{\rm X}|\psi)=\frac{1}{\sqrt{2\pi}\sigma}\exp\Big{[} \frac{-\big{(}\log \frac{L_{\rm X}}{\langle L_{\rm X}\rangle}\big{)}^2}{2\sigma^2}\Big{]},
\end{equation} where $\langle L_{\rm X} \rangle=c_{\rm X} \times \psi$ denotes the X-ray luminosity (measured in the 0.5-8.0 keV rest frame) that is expected from the observed SFR-$L_{\rm X}$ relation. The standard deviation $\sigma=0.4$ denotes the observed scatter in this relation \citep{Lehmer10,Mineo11}. \citet{Ranalli05} made very similar predictions for the XLFs, but instead of using star formation rate functions, they used galaxy luminosity functions in various bands. \citet{Ranalli05} also assumed a lognormal conditional probability functions for $P(\log L_{\rm X}|L_{\rm Y})$, where $L_{\rm Y}$ denotes the galaxy luminosity in some other band Y.

\subsection{Results}
\label{sec:resultsusxb}

Our fiducial model assumes $\Gamma=2.0$ (which corresponds to $\Gamma_{\rm pk}$, see above), $s_{\rm max}=2.4 \times 10^{-18}$ erg s$^{-1}$ cm$^{-2}$ (close to the middle of the range for $s_{\rm min}$ that was quoted in \S~\ref{sec:sxb}) , $\psi_{\rm min}=10^{-3}M_{\odot}$ yr$^{-1}$, and $z_{\rm max}=10$. As mentioned previously, we explore two choices for extrapolating the star formation rate function with redshift, which likely bracket the range of physically plausible models. When we evolve $\psi^*(z)$, but keep $\Phi^*$ fixed, we find that our fiducial model gives $S_{\rm X}=2.4\times 10^{-13}$ erg s$^{-1}$ deg$^{-2}$ cm$^{-2}$. On the other hand, when we evolve $\Phi^*(z)$, but keep $\psi^*$ fixed, we obtain $S_{\rm X}=2.6\times 10^{-13}$ erg s$^{-1}$ deg$^{-2}$ cm$^{-2}$. The fact that this difference is small is encouraging, and suggests that our ignorance of the star formation rate function at $z>0$ does not introduce major uncertainties into our calculations. We have verified that the second model generally yields slightly higher values for $S_{\rm X}$. To be conservative, we focus on the first model in the reminder of this paper.

Figure~\ref{fig:fig1} has four panels, each of which shows the median unresolved soft X-ray background ({\it black solid lines}), and its 68\% confidence levels (the {\it gray region}, bounded by {\it black dashed lines}) from Hickox \& Markevich (2007b). The {\it red solid bands} show the contribution from galaxies whose individual soft X-ray flux (1-2 keV observed frame) is less than $s_{\rm max}$. 

\begin{itemize}

\item In the {\it upper left panel} we plot $S_{\rm X}$ as a function
of $\psi_{\rm min}$. We find that $S_{\rm X}$ depends only weakly on
$\psi_{\rm min}$. That is, very faint galaxies do not contribute
significantly to $S_{\rm X}$. This is because the majority of the
contribution to $S_{\rm X}$ comes from galaxies at $z<2$ (see below,
and Fig.~\ref{fig:fig0}), where the `faint' end of the star formation
function increases as $\propto \psi^{-1.5}$, and the overall star
formation rate density is dominated by galaxies that are forming stars
at a rate close to $\psi^*$.\\

\item The {\it upper right panel} shows $S_{\rm X}$ as a function of $\Gamma$. We find that $S_{\rm X}$ decreases with $\Gamma$. As most of the contribution to $S_{\rm X}$ comes from galaxies at $z<2$, we are most sensitive to the X-ray emissivity of star forming galaxies at $E=[1-2]\times(1+z)<[3-6]$ keV (restframe). For fixed $L_{\rm X}$, increasing $\Gamma$ reduces the fraction of the emitted flux at these `higher' energies for steeper spectra (i.e. most of the energy lies near $E=0.5$ keV for the steepest spectra), which results in a decrease in $S_{\rm X}$.\\

\item The {\it lower left panel} shows $S_{\rm X}$ as a function of $s_{\rm max}$. The {\it vertical dotted line} shows the detection threshold in the {\it Chandra Deep Field-North} \citep{A03,HM07}. The {\it gray region} here brackets the effective minimum X-ray flux $s_{\rm min}$ that is probed by stacking X-ray undetected HST sources (see \S~\ref{sec:sxb}). Clearly, $S_{\rm x}$ depends only weakly on $s_{\rm max}$, unless $s_{\rm max}\lsim 10^{-18}$ erg s$^{-1}$ cm$^{-2}$. This weak dependence on $s_{\rm max}$ at larger fluxes can be easily understood: most of the contribution to $S_{\rm X}$ comes from $z\lsim 2$ (see below). For galaxies at $z=1$, $s_{\rm max}=2.4 \times 10^{-18}$ erg s$^{-1}$ cm$^{2}$ corresponds to $L_{\rm X}=4\pi d^2_{\rm L}(z) s_{\rm max}/\mathcal{L}_{\rm X}(z,\Gamma)=2.9\times 10^{40}$ erg s$^{-1}$, which requires $\psi=9.7 M_{\odot}$ yr$^{-1}$, which is close to $\psi^*$. We are therefore practically integrating over the full UV-luminosity function. Boosting $s_{\rm max}$ therefore barely increases $S_{\rm X}$ further.\\

\item The {\it lower right panel} shows $S_{\rm X}$ as a function of $z_{\rm max}$. This plots shows that $S_{\rm X}$ evolves most up to $z_{\rm max}\sim 2$, and barely when $z_{\rm max}\gsim 4$. That is, galaxies at higher redshift barely contribute to $S_{\rm X}$ as we also showed in Figure~\ref{fig:fig0} (unless the conversion factor $c_{\rm X}$ between $L_{\rm X}$ and SFR changes with redshift, see \S~\ref{sec:lxsfr}).

\end{itemize}

The main point of Figure~\ref{fig:fig1} is that it shows that {\it star forming galaxies that are too faint to be detected as individual X-ray sources, can account for the full unresolved SXB}, and that this statement is insensitive to details in the model when $\Gamma \lsim 2$ over the energy range 1-6 keV. This last requirement is very reasonable, given that the mean observed $\langle \Gamma \rangle =1.7$ for the majority of ULXs \citep{Sw04}.

\section{Constraining the Redshift Evolution of the SFR-$L_{\rm X}$ Relation}
\label{sec:lxsfr}

\subsection{Constraints from the SXB}
\label{sec:usxb}
\begin{figure}
\vbox{\centerline{\epsfig{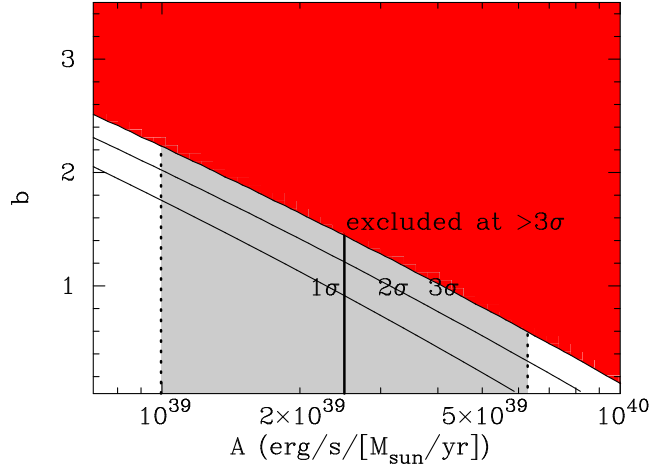}}}
\caption[]{The unresolved SXB constraints on the parameters $A$ and
$b$ for a redshift evolution parametrization of the form $c_{\rm X}
\equiv \frac{L_{\rm X}}{{\rm SFR}}=A(1+z)^b$. Models that lie in the
{\it red region} saturate the unresolved SXB at $> 3 \sigma$ (see
\S~\ref{sec:usxb}). The {\it grey region} denotes the value for
$c_{\rm X}(z=0) \equiv A$ derived by \citet{Mineo11} (the {\it solid
vertical line} denotes their best-fit value).  For this value of $A$,
$b \lsim 1.3$ (at $2\sigma$).}
\label{fig:con}
\end{figure}
\begin{figure*}
\vbox{\centerline{\epsfig{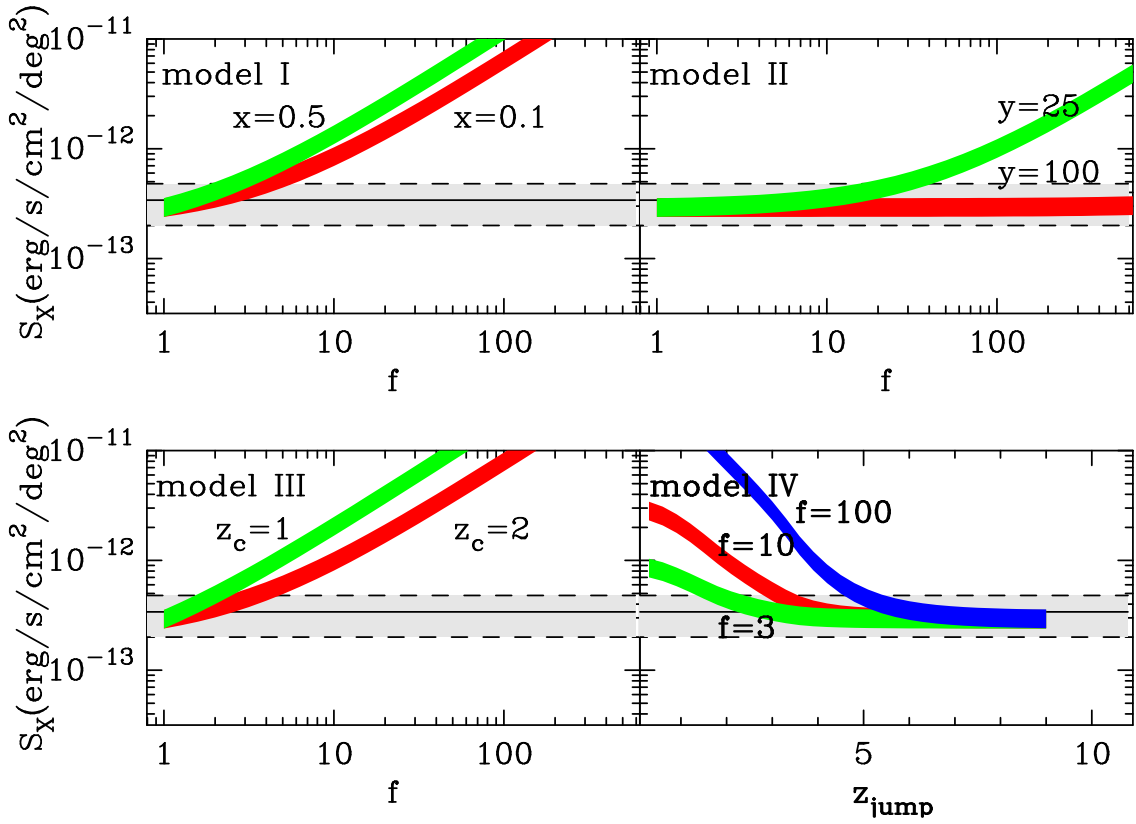}}}
\caption[]{Same as Figure~\ref{fig:fig1}, but we modify our fiducial model such that the X-ray emission from star forming galaxies is boosted by a factor of $c_{\rm X}\rightarrow f\times c_{\rm X}$ (i.e. $L_{\rm X}\rightarrow f\times L_{\rm X}$) when SFR$<x\hs M_{\odot}$ yr$^{-1}$ ({\it top left panel}), SFR$>y\hs M_{\odot}$ yr$^{-1}$ ({\it top right panel}), and z$>z_{\rm c}$  ({\it lower left panel}) as a function of $f$. The {\it lower right panel} shows a model where we boost $c_{\rm X}$ by a factor of 100 ({\it upper blue band}), 10 ({\it middle red band}) and 3 ({\it lower green band}) at redshift $z_{\rm jump}$ as a function of $z_{\rm jump}$. This figure shows that even small boosts (i.e. $f \sim$ a few) for SFR$<0.1 M_{\odot}$ yr$^{-1}$ or $z \geq 2$ violates the constraints posed by the unresolved SXB. It also shows that boosting $c_{\rm X}$ at the large SFR end barely affects $S_{\rm X}$. While large 'jumps' in $c_{\rm X}$ are not allowed if these occur at low redshift (i.e. $z\lsim 3$), the unresolved SXB cannot constrain large jumps at $z>5$.}
\label{fig:fig2}
\end{figure*}

Since our fiducial model already saturates the unresolved SXB, we can
ask what constraints we can set on the redshift-dependence of the
$L_{\rm X}$-SFR relation\footnote{As we already stated in \S~\ref{sec:intro}, there are several additional candidate sources which contribute to the unresolved SXB (see \S~\ref{sec:conc} for a summary). To constrain the evolution of $c_{\rm X}$ we will conservatively allow the entire unresolved SXB to be produced by HMXBs (and the hot ISM) in X-ray faint star-forming galaxies. Obviously, this results in upper limits on possible evolution in $c_{\rm X}$.}. Clearly, the SXB can only put constraints on
models in which $c_{\rm X}$ increases with redshift. We consider
models for which
\begin{equation}
\frac{L_{\rm X}}{{\rm SFR}}=c_{\rm X}\equiv A(1+z)^{b},
\label{eq:cxz}
\end{equation} 
with $b \geq 0$, and investigate the constraints that the SXB places
on the parameters $A$ and $b$. We compute $S_{\rm X}$ (see
Eq.~\ref{eq:sxfaint}) on a grid of models which cover a range of $A$
and $b$. Figure~\ref{fig:con} shows how many $\sigma(=1.4\times
10^{-13}$ erg s$^{-1}$ cm$^{-2}$ deg$^{-2}$) above the median observed
unresolved SXB, $S_{\rm obs}=3.4\times 10^{-13}$ erg s$^{-1}$
cm$^{-2}$ deg$^{-2}$, the models lie. Models that lie above the
uppermost {\it solid line}, indicated by the label `3-$\sigma$' result
in $S_{\rm X}-S_{\rm obs}>3\sigma$. That is, these models
significantly overproduce the unresolved SXB and are practically ruled
out. The {\it light grey region} bounded by the {\it dotted vertical
lines} indicates the $68\%$ confidence region for the observed value
for $c_{\rm X}$ by \citet{Mineo11}, where we assumed 0.4 dex
uncertainty on $c_{\rm X}$. Their best-fit value for $c_{\rm X}$ is
indicated by the {\it solid vertical line}. Figure~\ref{fig:con} shows
that for this value of $A$, $b \lsim 1.6$ at $\gsim 3 \sigma$. If we
marginalize\footnote{We obtain this marginalized upper limit
$\mathcal{U}(b)=\int_{A_{\rm min}}^{A_{\rm max}}dA \hs
\mathcal{U}(b|A)P_{\rm prior}(A)$. Here, $\mathcal{U}(b|A)$ denotes
the upper limit on $b$ given $A$, and $P_{\rm prior}(A)$ denotes our
prior on the probability density function for $A$.  Since $A\equiv
c_{\rm X}(z=0)$, we took $P_{\rm prior}(A)$ to be a lognormal
distribution with $\langle \log c_{\rm X} \rangle=39.4$, and
$\sigma=0.4$ \citep{Mineo11}. Finally, we adopted $A_{\rm
min}=10^{38.5}\hs {\rm erg} \hs{\rm s}^{-1}[M_{\odot}\hs{\rm
yr}^{-1}]^{-1}$, and $A_{\rm max}=10^{40.5}\hs{\rm erg} \hs{\rm
s}^{-1}[M_{\odot}\hs{\rm yr}^{-1}]^{-1}$.} over $A$, then we also get
$b \lsim 1.6$.

We also investigate changes in the SFR-$L_{\rm X}$ relation of the
form
\begin{eqnarray}
c_{\rm X}\rightarrow f\times c_{\rm X}\left\{ \begin{array}{ll}
     \  {\rm model \hs I} & {\rm SFR} \leq x \hs M_{\odot} \hs {\rm yr}^{-1};\\
         \ {\rm model \hs II} & {\rm SFR} \geq y \hs  M_{\odot} \hs {\rm yr}^{-1}\\
         \ {\rm models \hs III} \& {\rm IV} & z \geq z_{\rm c} \end{array} 
\right.
\end{eqnarray}
The {\it upper left panel} of Figure~\ref{fig:fig2} shows $S_{\rm X}$
as a function of $f$ for model I for $x=0.1$ ({\it red band}) and
$x=0.5$ ({\it green band}). This plot shows that $f\gsim 3-5$ for
$x=0.1$ is at odds with the unresolved SXB. This can be understood
from the {\it top left panel} of Figure~\ref{fig:fig1}, which shows
that adopting $\psi_{\rm min}=0.1 M_{\odot}$ yr$^{-1}$ results in $S_{\rm X}\sim 1.9 \times 10^{-13}$ erg s$^{-1}$ cm$^{-2}$ deg$^{-2}$, which corresponds to a reduction of $\sim 25\%$. Hence, galaxies with 0.001$<$$\frac{{\rm SFR}}{M_{\odot}\hs {\rm yr}^{-1}}<0.1$ contribute $\sim 25\%$ to $S_{\rm X}$ (for
$f=1$). Boosting their contribution by a factor $f \gsim 3-5$ (the
exact number depends on the precise fiducial choice for $c_{\rm X}$)
causes $S_{\rm X}$ to exceed the unresolved SXB. 

The {\it upper right
panel} shows that the soft XRB only allows constraints to be put on
$f\gsim 10$, and only if $y \lsim 25$. Strong constraints on $f$ are
not possible for large $\psi$. This is because galaxies that are
forming stars at a rate $\psi \gsim 100 M_{\odot}$ yr$^{-1}$ are deep
in the exponential tail of the star formation function. As a result of
their small number density, they barely contribute anything to $S_{\rm
X}$. 

The {\it lower left panel} shows that boosting $c_{\rm X}$ at $z
\geq 2$ by factors greater than $f\gsim 3-5$ is again at odds with the
unresolved SXB. This is because galaxies at $z>2$ contribute
noticeably to $S_{\rm X}$ for our fiducial choice of $c_{\rm
X}$. However, the unresolved SXB cannot place tight limits (yet) on
$c_{\rm X}$ at very high redshifts. This is illustrated in the {\it
lower right panel} where we show $S_{\rm x}$ for model IV: once
$z_{\rm jump}\gsim 5$, boosting $c_{\rm X}$ by as much as $\sim 100$
has little impact on $S_{\rm X}$.

\subsection{Constraints from the Galaxy XLFs?}
\label{sec:xnum}
As part of our analysis, we compute theoretical galaxy X-Ray
luminosity functions (XLFs) using equation~(\ref{eq:xlf}). Observed
galaxy XLFs have been presented, for example, by
\citet{Norman04,G05,G06,T08}. \citet{T08} present galaxy XLF for late-type
(i.e. star forming) galaxies in two redshifts bins: the first bin
contains galaxies with $0 < z< 0.4$ ($z_{\rm med}=0.14$), and the
second bin contains galaxies with $0.4 \leq z < 1.4$ ($z_{\rm
med}=0.68$). Note that \citet{T08} derive X-Ray luminosities in the
0.5-2.0 keV band, and we properly K-correct our models into their
band.

\begin{figure}
\vbox{\centerline{\epsfig{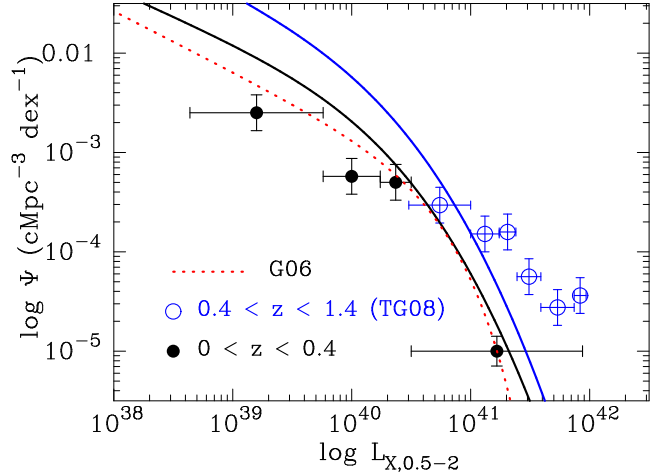}}}
\caption[]{The observed X-Ray luminosity functions (XLFs) of galaxies from Tzanavaris \& Georgantopoulos (2008, {\it data points}) and Georgakakis et al (2006, the {\it red dotted line} shows their best-fit Schechter function) are compared to our fiducial model $c_{\rm X}=A(1+z)^b$ with $A=2.6\times 10^{39}$ erg s$^{-1}$ [$M_{\odot}$ yr$^{-1}$]$^{-1}$, and $b=0.0$. At $z\leq 0.4$ our model provides a good fit at the two brightest X-ray luminosities, but overpredicts the number density of fainter sources by a factor of $\sim 2-3$. The $z \leq 0.4$ XLF prefers $c_{\rm x}$ to be lower by $\sim 0.2$ dex, within the dispersion found by \citet{Mineo11}. At higher redshift, our model significantly underpredicts the number density at $L_{\rm X,0.5-2.0} \gsim 10^{41}$ erg s$^{-1}$. This suggests that either $b>0$, or that $c_{\rm X}$ increases at high $\psi$ (see text).}
\label{fig:norman}
\end{figure}

In Figure~\ref{fig:norman} the {\it data points} show the observed
X-Ray luminosity functions (XLFs) from Tzanavaris \& Georgantopoulos
(2008), while the {\it red dotted line} shows the best-fit Schechter
function derived by Georgakakis et al (2006). The {\it solid lines}
show predictions for our fiducial model $c_{\rm X}=A(1+z)^b$ with
$A=2.6\times 10^{39}$ erg s$^{-1}$ [$M_{\odot}$ yr$^{-1}$]$^{-1}$, and
$b=0.0$. At $z\leq 0.4$ our model provides a good fit at the two
brightest X-ray luminosities, but overpredicts the number density of
fainter sources by a factor of $\sim 2-3$. We obtain a better fit if
we lower $A=1.4\times 10^{39}$ erg s$^{-1}/[M_{\odot}$yr$^{-1}]$,
which corresponds to $\sim 0.2$ dex, and thus lies within the
dispersion that was found by \citet{Mineo11}. This lower
value of $A$ corresponds to almost exactly the value quoted by Lehmer
et al (2010, their $\beta$, although these authors measured $L_{\rm X}$ in the 2.0-10.0 keV range). At higher redshift, our model
significantly underpredicts the number density at $L_{\rm X,0.5-2.0}
\gsim 10^{41}$ erg s$^{-1}$. This may suggest that either that the XLFs
(strongly) favor $c_{\rm X}$ to increase towards higher redshift,
and/or SFR $\gsim$ a few tens of $M_{\odot}$ yr$^{-1}$ (see below). Alternatively, the observed XLFs of star forming galaxies are contaminated by low luminosity AGN, which are difficult to identify at these X-ray luminosities.

Our predicted XLFs agree quite well with previous predictions by \citet{Ranalli05}, for the model in which they assume that the redshift evolution of the luminosity functions is solely the result of evolution in the number density of galaxies (this is referred to as `density evolution').  Their model also underpredicts the number density of luminous X-ray sources at higher redshift. \citet{Ranalli05} found that a better fit to the high-redshift data is obtained for a model in which solely the luminosity of galaxies evolves  (`luminosity evolution') as $\propto (1+z)^{2.7-3.4}$. Our work also indicates that evolution in the number density of star forming galaxies is not enough to explain the observed redshift evolution of XLFs, and that some additional luminosity evolution is preferred.

\begin{figure}
\vbox{\centerline{\epsfig{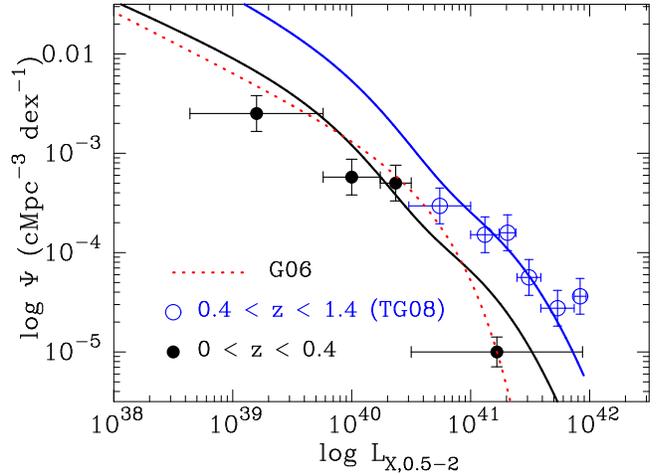}}}
\caption[]{Same as Figure~\ref{fig:norman}, except that here we
improved the fit to the observations by forcing $c_{\rm X}$ to evolve
with redshift and star formation rate. The model now assumes $c_{\rm
X}=A(1+z)^b$ with $A=1.4\times 10^{39}$ erg s$^{-1}$ [$M_{\odot}$
yr$^{-1}$]$^{-1}$, and $b=1.0$. Furthermore, we boosted $c_{\rm X}$ by
a factor of $\sim 3$ at SFR$\gsim 15 M_{\odot}$ yr$^{-1}$. This plot
illustrates that the XLFs favor an increase of $c_{\rm X}$ with
redshift and/or at larger SFR.}
\label{fig:norman2}
\end{figure}

It is possible to compute the likelihood
$\mathcal{L}(A,b)=\exp[-0.5\chi^2]$ by fitting to the observed XLFs
for any combination of $A$ and $b$ describing the redshift evolution
of $c_{\rm X}$ (see Eq.~\ref{eq:cxz}). However, we found that this
formal fit is dominated by the two lowest luminosity data points at
$z\leq 0.4$, which lie significantly below the XLF that was derived by
\citet{G06}. Furthermore, we found that this procedure also depends
somewhat on the assumed redshift evolution of the star formation rate
function. We therefore do not pursue a detailed statistical analysis
on constraining the redshift evolution of $c_{\rm X}$ with the
XLFs. Instead, we show in Figure~\ref{fig:norman2} {\it an example} of
a model where $c_{\rm X}$ increases both with redshift and at high
SFR: this model with $A=1.4\times 10^{39}$ erg s$^{-1}$ [$M_{\odot}$
yr$^{-1}$]$^{-1}$, $b=1.0$, and $c_{\rm X} \rightarrow 3c_{\rm X}$ at
SFR$\geq 15 M_{\odot}$ yr$^{-1}$ fits the observed XLFs much
better. The value $b=1.0$ is consistent with the X-ray background
constraint (see Fig.~\ref{fig:con}), and boosting $c_{\rm X}$ by a
factor of $3$ at SFR$\geq 15 M_{\odot}$ yr$^{-1}$ is also consistent
with the soft XRB (see the {\it top right panel} of
Fig.~\ref{fig:fig2}). However, boosting $c_{\rm X}$ by a factor of $3$
at SFR$\geq 15 M_{\odot}$ yr$^{-1}$ appears inconsistent with direct
constraints on the $L_{\rm X}$-SFR relation. In
Figure~\ref{fig:norman2} we used $c_{\rm X}=4.2(1+z)\times 10^{39}$
erg s$^{-1}$ [$M_{\odot}$ yr$^{-1}$]$^{-1}$ at SFR$\gsim 15 M_{\odot}$
yr$^{-1}$ which translates to $c_{\rm X}=8.4\times 10^{39}$ erg
s$^{-1}$ [$M_{\odot}$ yr$^{-1}$]$^{-1}$ at $z=1$, while
\citet{Mineo11} found $c_{\rm X}=3.7\times 10^{39}$ erg s$^{-1}$
[$M_{\odot}$ yr$^{-1}$]$^{-1}$ for their sample of unresolved, high
SFR, sources at $z\sim 0.2-1.2$. However, the sample of high-SFR,
high-z galaxies that was studied by \citet{Mineo11} is rather limited,
and if the dispersion around this mean quantity is also 0.4 dex, then
our model may be consistent with the observed dispersion around this
value out to $z\sim 1$.

\section{Discussion \& Conclusions}
\label{sec:conc}

Observations have established that a correlation exists between the star formation rate of galaxies and their X-ray luminosity (measured over the range E=0.5-8 keV, e.g. Ranalli et al. 2003, Grimm et al. 2003, Gilfanov et al. 2004, Lehmer et al. 2010, Mineo et al. 2010). This `$L_{\rm X}$-SFR relation' encodes a wealth of information on various astrophysical processes, and strongly affects the thermal evolution of the intergalactic medium during the early stages of the epoch of reionization. Existing observations have only been able to probe this relation in nearby galaxies, and while theoretically there are good reasons to suspect that $c_{\rm x}\equiv L_{\rm X}/$SFR increases towards higher redshifts, observational constraints are virtually non-existent.

In this paper, we have investigated whether it is possible to put any
constraints on the evolution of $c_{\rm x}$ with either redshift
and/or towards high/low star formation rates than probed by existing
observations of individual galaxies. As part of our analysis, we have
computed that the observed `local' relation, when combined with the
most observational constraints on the redshift-evolution of the star
formation rate density of our Universe, implies that star forming
galaxies contribute $\sim 5-15\%$ of both the soft and $\sim 1-20\%$
of the hard X-ray backgrounds (see \S~\ref{sec:totsfr}). The ranges
that we quoted are for a range of photon index $1< \Gamma <3$. The
observed $\Gamma$ of ULX spectra in the Chandra bands is described by
a distribution with a mean of $\langle \Gamma \rangle=1.7$, and a mode
of $\Gamma_{\rm pk}\sim 2.0$ \citep{Sw04}. For the range
$\Gamma=1.7-2.0$ the fractional contribution of star forming galaxies
to the soft X-ray background is $\sim 8-13 \%$. The contribution to
the HXB remains uncertain, mostly because of a more uncertain
K-correction at the corresponding high energies \citep[also see][for earlier calculations of the contribution of star forming galaxies to the CXBs]{TL96,NA00}.

We have then taken the most recent observational constraints on the
star formation rate function, which gives the comoving number density
of star forming galaxies as a function of their star formation rate
[denoted by $n(\psi,z)$], and computed what the contribution of `X-ray
faint' star forming galaxies to the soft X-ray background (SXB,
corresponding to 1-2 keV in the observed frame) is. We found that
galaxies whose individual observed flux is $s \leq s_{\rm max}=2.4
\times 10^{-18}$ erg s$^{-1}$ cm$^{-2}$ between 1-2 keV, i.e.  more
than an order of magnitude fainter than the detection threshold in the
{\it Chandra Deep Field-North} (see \S~\ref{sec:sxb}), can fully
account for the unresolved SXB. This statement is insensitive to
details in the model as longs as the photon index, averaged over the
entire population of X-ray emitting star forming galaxies, is $\Gamma
\lsim 2$, which corresponds to a very reasonable range given the
existing observational constraints on this parameter
(\S~\ref{sec:usfr}).

Motivated by our result that X-ray faint star forming galaxies can
fully account for the unresolved SXB, we put constraints on the
redshift evolution of the parameter $c_{\rm X}$. When we parametrize
the redshift evolution as $c_{\rm X}=A(1+z)^b$, we found that the
unresolved SXB requires that $b \lsim 1.6$ ($3 \sigma$). We have also
ruled out models in which $c_{\rm X}$ is boosted by a factor of
$f\gsim 2-5$ at $z \geq 1-2$ and/or SFR $\leq 0.1-0.5 M_{\odot}$ yr$^{-1}$, as they overproduce the unresolved SXB (see {\it
left panels} of Fig.~\ref{fig:fig2}). We have found indications in the
observed X-ray luminosity functions (XLFs) of star forming galaxies
that $c_{\rm X}$ is increasing towards higher redshifts and/or higher
star formation rates , but caution that this may indicate the presence of unidentified low luminosity AGN.  The unresolved SXB allows for larger changes in
$c_{\rm X}$ at large values for SFR (see the {\it top right panel} of
Fig.~\ref{fig:fig2}). Finally, we also found that the SXB puts weak
constraints on possible strong evolution ($f \sim 100$) at $z >5$ (see
the {\it lower right panel} of Fig.~\ref{fig:fig2})\footnote{\citet{Treister11} stacked 197 HST detected candidate $z\sim 6$ galaxies and found significant X-ray detections in both the soft (0.5-2.0 keV) and hard (2.0-8.0 keV) bands (but see Cowie et al. 2011 and Willott 2011). They derive an average rest frame 2-10 keV luminosity of $L_{\rm X,2-10}=6.8 \times 10^{42}$ erg s$^{-1}$, which they associate with obscured AGN. We can use this detection to place an upper limit on $c_{\rm X}$ at $z=6$. The mean star formation rate --averaged over the UV-luminosity function in the range $-21.5 < M_{\rm UV} < -18.0$, and not corrected for dust-- is $\sim 2 M_{\odot}$ yr$^{-1}$. The stacked X-ray detection therefore puts an upper limit on the boost $f\leq L_{\rm X,2-10}/1.5c_{\rm X}\sim 10^3$. We verified that such a boost at $z_{\rm jump}\leq 6.2$ is ruled out at $95\%$ CL. }.

There are many other undetected candidate sources which likely also
contribute to the unresolved SXB. These are briefly summarized below
\citep[see][for a more detailed summary]{D04}:
\begin{itemize}

\item Observed AGN account for $\sim 80\%$ of the SXB. It would be
highly unlikely that fainter AGN-- i.e. those are too faint to be
detected as individual X-ray sources--do not provide a significant
contribution to the unresolved SXB.

\item  Our attention has focused on HMXBs, but low mass X-Ray binaries
(LMXBs)-in which the primary has a mass $\lsim 5 M_{\odot}$, dominate
the X-ray luminosity of galaxies for which the specific star formation
is sSFR$\lsim 10^{-10}$ yr$^{-1}$ \citep{Gilfanov04,Lehmer10}. LMXBs
give rise to a correlation between X-ray luminosity and total stellar
mass, $M_*$, which is $L_{\rm X,LMXB}\sim 9\times 10^{28} M_{*}$ erg
s$^{-1}$ \citep{G04,Lehmer10}. In Appendix~\ref{app:lmxb} we repeat the calculation of \S~\ref{sec:usfr} and replace the star formation rate function
$n(\psi,z)$ with the stellar mass function $n(M_*,z)$, and appropriately replace $P(\log L_{\rm
X}|\psi)$ with $P(\log L_{\rm X}|M_*)$. We found that faint
`quiescent' galaxies contribute about an order of magnitude less to the SXB than faint star-forming galaxies.

\item Thomson scattering of X-rays emitted predominantly by high-redshift sources can cause 1.0-1.7\% of the SXB to be in a truly diffuse form \citep{Soltan03}. Similarly, intergalactic dust could scatter X-rays by small angles into diffuse halos that are too faint to be detected individually \citep{Petric06,DL09}. 

\item \citet{Wu01} computed that clusters and groups of galaxies possibly contribute as much as $\sim 10\%$ of the total SXB.

\item A (hypothetical) population of `miniquasars' powered by intermediate mass black holes may have contributed to ionizing and heating the IGM \citep{Madau04,Ricotti04}. These miniquasars would emit hard X-ray photons that could contribute significantly to the soft and hard X-ray backgrounds \citep{Ricotti04,D04}.

\end{itemize}

The likely existence of these additional contributors to the
unresolved SXB implies that our constraints are conservative, and that
actual limits on the redshift evolution of $c_{\rm X}$ should be
tighter.

 After our paper was submitted, Cowie et al. (2011) compared the average X-ray fluxes (obtained by a stacking analysis) and  restframe UV flux densities of sources with known redshifts in the 4 Ms exposure of the CDF-S field.  Cowie et al. (2011) showed that this ratio --after an extinction correction-- was consistent with the local $L_{\rm X}$--SFR relation up to $z\sim 4$. The stacking of many source allowed Cowie et al. (2011) to probe down to $s \sim s_{\rm max}/4 =5 \times 10^{-19}$ erg s$^{-1}$ cm$^{-2}$, which translates to a luminosity of $L_{\rm X}\sim 0.3-4 \times 10^{40}$ erg s$^{-1}$ at $z=1-3$. \citet{Cowie11} therefore probe the redshift-evolution of $c_{\rm X}$ at SFR$\gsim 1-10 M_{\odot}$ yr$^{-1}$ at these redshifts (depending on $z$ and $c_{\rm X}$). For comparison, we have shown that the SXB allows for constraints at lower SFR, but that the SXB becomes less sensitive to changes in $c_{\rm X}$ at $z \gsim 2$. Our results in combination with those of Cowie et al. (2011) thus provide stronger constraints on the allowed redshift evolution of $c_{\rm X}$. Interestingly,  a non-evolution in $c_{\rm X}$ with redshift appears at odds with the observed redshift evolution in the XLFs (unless these are contaminated by low luminosity AGN, see above).

Constraints on the redshift evolution of the  $L_{\rm X}$--SFR relation will be helpful in pinning down the
astrophysics that is driving the $L_{\rm X}$-SFR relation, and may
eventually give us new insights into the X-ray emissivity of the first
galaxies which plays a crucial role in determining the thermal history
IGM during the dark ages \citep{Mirabel11}.

{ \bf Acknowledgements} We thank David Schiminovich, Tassos Fragos,
and Stefano Mineo for helpful discussions. MD thanks Harvard's
Institute for Theory and Computation (ITC) for its kind hospitality.
This work was supported in part by NSF grant AST-0907890 and NASA
grants NNX08AL43G and NNA09DB30A (for A.L.).

\appendix

\section{Assumed Redshift Evolution of the Star Formation Rate Function}
\label{app:dndpsi}

\subsection{The Integrated Star Formation Rate Density}
\begin{figure}
\vbox{\centerline{\epsfig{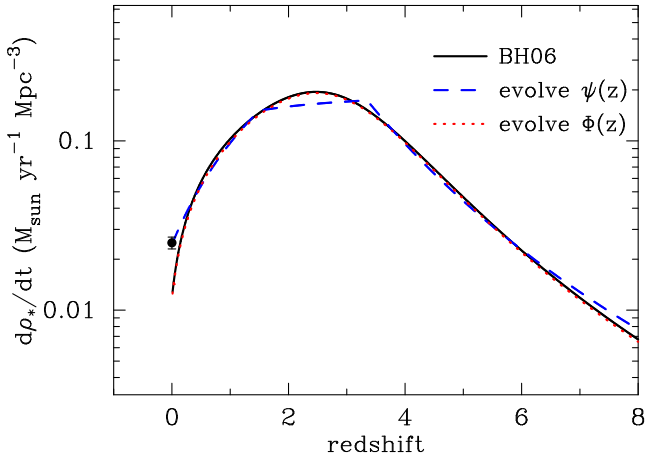}}}
\caption[]{This plot shows the star formation rate density (comoving) in the Universe, $\dot{\rho}_*(z)$. The {\it black solid line} shows  $\dot{\rho}_*(z)$ that has been derived by \citet{Hopkins06}. The data point at $z=0$ represents the more recent $z=0$ estimate by Bothwell et al. (2011). The {\it blue dashed line} ({\it red dotted line}) shows our model in which we evolve $\psi^*$ ($\Phi^*$) in our adopted star formation rate function with redshift, while keeping $\Phi^*$ ($\psi^*$) fixed. Both models clearly reproduce the `observed'  redshift evolution of the star formation rate density.}
\label{fig:sfrden}
\end{figure}
In our paper, we needed to assume the redshift evolution of the star
formation rate function (${dn}/{d\psi}$). We studied two models:
\begin{itemize}
\item In our first model, we evolved $\psi^*(z)$ to match the observed star formation rate density, but kept $\Phi^*$ fixed. We found that  the following redshift evolution of $\psi^*(z)$ provides a decent fit to observations:

\begin{eqnarray}
\frac{\psi^*(z)}{M_{\odot}\hs {\rm yr}^{-1}}=\left\{ \begin{array}{ll}
     \ 9.2(1+z)^{0.8}& z < z_1;\\
         \ 9.2(1+z_1)^{0.8} & z_1 \leq z < z_2\\
          \ 9.2(1+z_1)^{0.8}\Big{(}\frac{1+z_2}{1+z}\Big{)}^{2.2} &z \geq z_2 ,\end{array} 
\right. 
\end{eqnarray} where $z_1=1.5$ and $z_2=3.35$. The resulting integrated star formation rate density is shown as the {\it blue dashed line} in Figure~\ref{fig:sfrden}, which should be compared to that derived by \citet{Hopkins06}, which is shown as the {\it black solid line}.  Both curves clearly agree. Note that our adopted star formation rate function--which was compiled from the most recent data--results in a larger star formation rate density at $z=0$ ($\dot{\rho}_*=0.025M_{\odot}$ yr$^{-1}$ Mpc$^{-3}$, indicated by the {\it black filled circle} at $z=0$), than that inferred by \citet{Hopkins06}, which is $\dot{\rho}_{\rm BH06}(z=0)=ah=0.012 M_{\odot}$ yr$^{-1}$ Mpc$^{-3}$.

\item In our second  model, we evolved $\Phi^*(z)$ to match the observed star formation rate density, but kept $\psi^*$ fixed. We assumed the redshift evolution of $\Phi^*(z)$ was 

\begin{eqnarray}
\Phi^*(z)=\Phi^*(z=0)\times\frac{\dot{\rho}_{\rm
HB06}(z)}{\dot{\rho}_{\rm HB06}(z=0)}g(z),
\end{eqnarray}
where the function $g(z)\equiv \frac{1}{2}+\frac{1}{2}(1+z)^{-1}$
compensates for the fact that the faint-end slope of the star
formation rate function, $\alpha$, changes with redshift in our
model. The resulting integrated star formation rate density is shown
as the {\it red dotted line} in Figure~\ref{fig:sfrden}.

\end{itemize}

\section{Derivation of Eq~1}
\label{app:der}
Eq~\ref{eq:sx} plays a central role in our analysis. Here, we provide more details on its origin. The total observed flux $dS$ from a proper (i.e physical) cosmological volume element $dV_{\rm p}$ is $dS(z)=(1+z)^3 dV_{\rm p} \dot{\rho}_* \mathcal{L}(z,\Gamma)/4 \pi d_{\rm L}^2(z)$. The cosmological proper volume element $dV_{\rm p}$ can be written as $dV_{\rm p}=\frac{c}{H_0}\frac{dz}{(1+z)\mathcal{E}(z)}dA_{\rm p}$. We substitute $dA_{\rm p}=d^2_{\rm A}(z)d\Omega$, where $d_{\rm A}(z)$ denotes the angular diameter distance to redshift $z$. We finally get for the differential flux {\it per sterradian}

\begin{eqnarray}
\frac{dS(z)}{d\Omega}=\frac{c}{H_0}\frac{d^2_{\rm A}(z)\dot{\rho}_*(z)(1+z)^3\mathcal{L}(\Gamma,z)}{(1+z)\mathcal{E}(z) 4 \pi d_{\rm L}^2(z)}= \\ \nonumber
\frac{c}{4\pi H_0} \frac{\dot{\rho}_*(z)\mathcal{L}(\Gamma,z)dz}{\mathcal{E}(z)(1+z)^2},
\end{eqnarray} where we used that $d_{\rm A}(z)=(1+z)^{-2}d_{\rm L}(z)$. When we integrate over redshift and solid angle $\Delta \Omega$, we arrive at Eq~\ref{eq:sx}.

\section{Contribution of Quiescent Galaxies to the SXB}
\label{app:lmxb}
\begin{figure*}
\vbox{\centerline{\epsfig{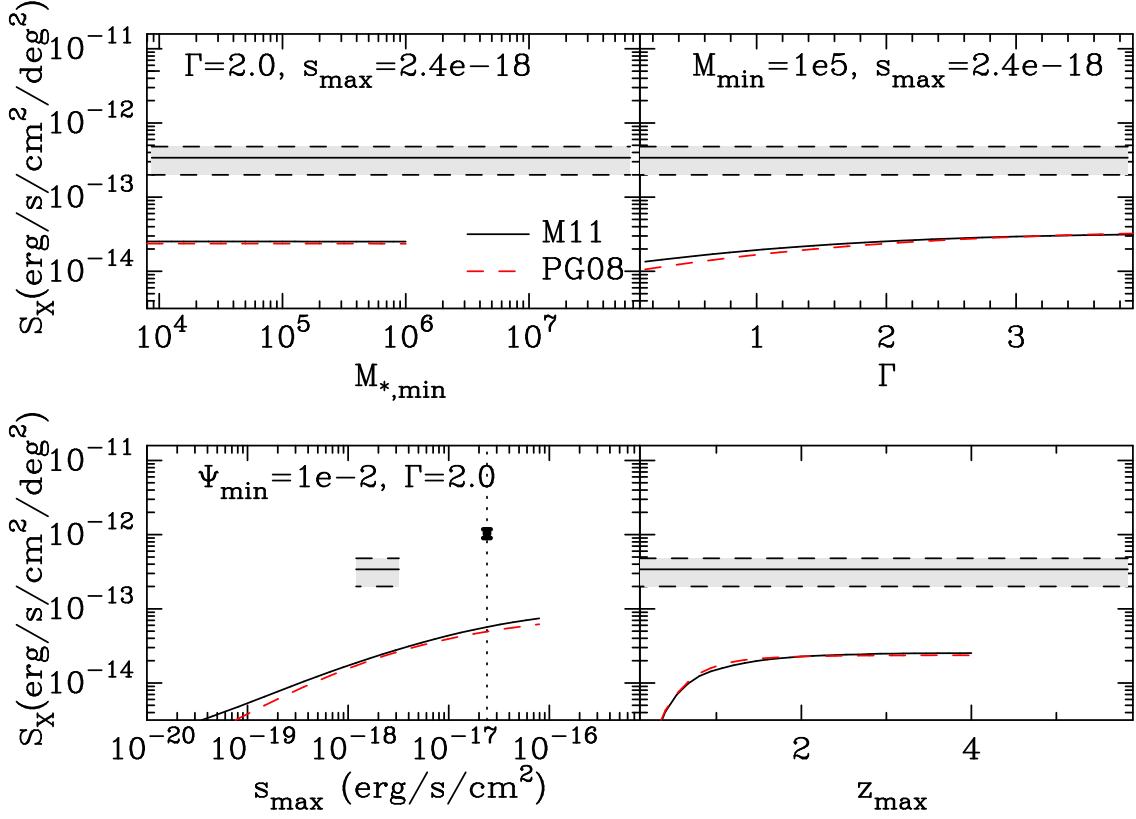}}}
\caption[]{Same as Figure~\ref{fig:fig1}, but for low mass X-ray binaries (LMXB), for which the cumulative luminosity scales linearly with the total stellar mass. We integrate the stellar mass functions down to some minimum mass $M_{\rm min}$. The {\it upper left panel} shows $S_{\rm X}$ as a function of $M_{\rm min}$. Another difference with Figure~\ref{fig:fig1} is that we do not extend our calculation beyond $z_{\rm max}=4.0$ as the stellar mass functions are constrained poorly at these redshifts. The {\it black solid lines} [{\it red dashed lines}] show $S_{\rm X}$ if we adopt the stellar mass functions from Mortlock et al. (2011) [P{\'e}rez-Gonz{\'a}lez et al. 2008].  We find that typically, the contribution from LMXBs to the SXB lies about an order of magnitude below that of HMXBs.}
\label{fig:lmxb}
\end{figure*}

To compute the contribution of low mass X-ray binaries to the SXB we only need to modify Eq~\ref{eq:xlf} in two ways: (i) we replace the star formation rate function with the observed stellar mass function $n(M_*,z)$, and (ii) we replace $P(\log L_{\rm X} | \psi)$ with $P(\log L_{\rm X}|M_*)$. The goal of this appendix is to provide more details of the calculation, and to show that the conclusion that LMXBs contribute about an order of magnitude less to the SXB than HMXBs is robust to uncertainties in the modeling. 

Observed stellar mass functions can be described by Schechter functions, and observations have constrained the Schechter parameters out to $z\sim 4$ \citep[e.g.][]{PG08,M09,Kajisawa09,Mortlock11}. In Figure~\ref{fig:lmxb} we show results from calculations in which we adopted the parameters from P{\'e}rez-Gonz{\'a}lez et al. (2008, {\it red dashed lines}), and Mortlock et al. (2011, {\it black solid lines}). Both calculations agree well.

We assume that $P(\log L_{\rm X}|M_*)$ is given by a lognormal distribution, where $\langle L_{\rm X} \rangle \equiv \mathcal{C}_{\rm X} M_{*}$. Here, $\mathcal{C}_{\rm X}=8.0\pm 0.5 \times 10^{28}$ erg s$^{-1}$ $M^{-1}_{\odot}$ \citep{Gilfanov04}.  The scatter in this relation is not given, and for simplicity we have adopted $\sigma_{\rm 1}=0.4$, but note that our results can simply be rescaled by a factor of $\exp\big{(}\frac{1}{2} {\rm ln}^2 10[\sigma_{\rm 2}^2-\sigma^2_{\rm 1}]\big{)}$ to obtain predictions for any $\sigma_{\rm 2}$. The last difference with the calculation presented in the main paper is that $L_{\rm X,LMXB}$ is measured in the restframe E$=2-10$ keV band \citep{Lehmer10}.

Figure~\ref{fig:lmxb} presents results from our calculations in a way that is identical to Figure~\ref{fig:fig1} of the main paper. The main differences are: ({\it i}) the {\it upper left panel} shows $S_{\rm X}$ as a function of minimum stellar mass (instead of minimum star formation rate), and ({\it ii}) our calculations extend only out to $z_{\rm max}=4.0$, as the observed stellar mass functions become uncertain there. The general trends in this figure are similar to those in Figure~\ref{fig:fig1}, except the dependence of $S_{\rm X}$ on $\Gamma$. This different dependence results from the fact that $L_{\rm X,LMXB}$ is measured in the 2-10 keV band (compared to 0.5-8.0 keV for HMXBs) which introduces different K-corrections. Generally, we find that LMXBs produce $S_{\rm X} \lsim 3 \times 10^{-14}$ erg s$^{-1}$ cm$^{-2}$ deg$^{-2}$, which is $\sim 10\%$ of the amount contributed by HMXBs.  

\label{lastpage}
\end{document}